\documentclass{aa}
\usepackage{graphicx}
\usepackage{natbib}
\usepackage{txfonts}
\bibpunct{(}{)}{;}{a}{}{,}

\begin{document}

\title{{Detailed chemical composition of Galactic Cepheids.}
\subtitle{A determination of the Galactic abundance gradient in the 8-12 kpc region.}
\thanks {Based on observations made with the 1.52m ESO Telescope at La Silla, Chile.}
}

\author{
B. Lemasle  \inst{1,2},
P. Fran\c cois  \inst{3,2},
G. Bono  \inst{4,5},
M. Mottini  \inst{5},
F. Primas \inst{5}, 
M. Romaniello  \inst{5} 
         }

\offprints {B. Lemasle}

\institute { 
             Universit\'e de Picardie Jules Verne, Facult\'e des Sciences, 33 rue Saint-Leu , 
	      80039 Amiens Cedex 1
          \and
           Observatoire de Paris-Meudon, GEPI, F-92195 Meudon Cedex, France \\
    \email {Bertrand.Lemasle@etud.u-picardie.fr} 
          \and
           European Southern Observatory (ESO), 3107 Alonso de Cordova, Vitacura, Casilla 19001, Santiago 19, Chile 
          \and
           INAF-Osservatorio Astronomico di Roma, via Frascati 33, I-00040 Monte Porzio Catone, Italy
          \and
           European Southern Observatory (ESO) , Karl-Schwarzschild-Strasse 2, D-85748 Garching bei Muenchen, Germany 
}
          
\authorrunning {Lemasle et al.}

\titlerunning {Detailed chemical composition of Galactic Cepheids.}

\date{Received xxx; accepted xxx}

\abstract
{}
{The recent introduction of high-resolution/large spectral-range spectrographs has provided 
the opportunity to investigate the chemical composition of classical Cepheids in detail. 
This paper focusses on new abundance determinations for iron and 6 light metals 
(O, Na, Mg, Al, Si, Ca) in 30 Galactic Cepheids. We also give a new estimate of the 
Galactic radial abundance gradient.}
{The stellar effective temperatures were determined using the method of line depth 
ratios, and the surface gravity and the microturbulent velocity v$_{t}$ by imposing the 
ionization balance between Fe I and Fe II with the help of curves of growth. Abundances 
were calculated with classical LTE atmosphere models.}
{Abundances were obtained with RMS accuracies of about 0.05-0.10 dex for Fe, and 
0.05-0.20 dex for the other elements. Cepheids in our sample have solar-like abundances, 
and current measurements agree quite well with previous determinations. We computed  
``single zone'' Galactic radial abundance gradients for the 8-12 kpc region and 
found a slope for iron of -0.061 dex kpc$^{-1}$.}
{}

\maketitle

%

\section{Introduction}

Galactic abundance gradients are one of the most important inputs for the numerical 
models of the chemical evolution of Galactic disks. Different stellar tracers have been 
used to estimate the gradient, namely planetary nebulae, giant stars, old open clusters, 
HII regions, and young, $B-$type main sequence stars. If the existence of the gradient 
is now widely accepted, its precise value still needs to be established. 

As a matter of fact, empirical estimates give quite steep gradients for iron ranging 
from -0.05 dex kpc$^{-1}$ to -0.07 dex kpc$^{-1}$: see, e.g., \citet{Frie02} who found 
a slope of -0.06 dex kpc$^{-1}$; \citet{Kov05}, who obtained a gradient of -0.06 dex kpc$^{-1}$; 
and \citet{Luck06}, who estimated a slope of -0.068 dex kpc$^{-1}$ using classical Cepheids 
as chemical tracers. However, much shallower slopes 
have also been estimated: -0.023 dex kpc$^{-1}$ by \citet{Twa97}, -0.044 dex kpc$^{-1}$ 
by \citet{And04}. And in a few cases the gradient was not detected: -0.017 dex kpc$^{-1}$ 
by \citet{Nees88} and -0.003 dex kpc$^{-1}$ by \citet{Kil94}. 
The broad range of empirical radial gradients also applies to the other light 
metals adopted in current investigations. Thorough reviews of gradient values for 
many elements can be found for example in \citet{And02a} and \citet{Chia01}.

Beyond an expected scatter due to measurement errors, these discrepancies between 
steep and shallow slopes of the Galactic radial abundance gradient might trace 
its evolution as a function of stellar ages. This working hypothesis has been suggested 
in several studies \citep{Frie02,Mac03} based either on planetary nebulae or on open clusters.
From this point of view, Cepheids are young Galactic objects that trace the present-day 
abundance gradient, and only a few elements (C, N, Na, and perhaps Mg and Al); \citep{Kov96} 
may have their abundances modified during the life of the Cepheids, while all the others should 
trace the abundance pattern as it was at the birth of the star. 

Compared to other methods, the use of classical Cepheids offerss many advantages because 
these stars
\begin{itemize}
	\item are excellent distance indicators;
	\item are luminous enough and provide the opportunity of determining the gradient 
over a significant fraction of the Galactic disk;
	\item allow the measurements of many elements that cannot be investigated with 
other stellar tracers;  
	\item show spectra with many absorption lines and, in turn, reliable abundance values. 
\end{itemize}

An extension of the available data is mandatory for constraining the slope of the Galactic 
abundance gradients on a more quantitative basis. This paper 
is the first in a series focussed on accurate and 
homogeneous abundance measurements for a good sample of Galactic Cepheids. 
The sample includes objects that are in common with previous studies, 
but current spectra typically present a higher spectral resolution and/or a 
wider spectral range.

%

\section{ Observations}

The observations were performed in 2002 at the 1,52m ESO telescope
in La Silla. We collected 38 high-resolution spectra of Galactic Cepheids 
with the fibre-fed Extended Range Optical Spectrograph (FEROS), with a 
resolving power of 48,000 and a spectral range spanning from 370 to 920 nm.
The spectra were reduced using the FEROS package within MIDAS. 
The S/N ratio at 600 nm ranges from 40 to 136 (see Table \ref{obslog}).

\begin {table*}[!ht]

\scriptsize 
\caption {Observations: date, exposure time, and S/N at 600 nm.} 

\begin {tabular}{lccr} 
\hline
\hline
 Object   &    date    &  exposure time (s) &   S/N  \\
 \hline 
 AD Pup   &  02/01/01  &         600        &    65  \\
 AH Vel   &  29/12/00  &         300        &   111  \\
 \ldots   &  31/12/00  &         100        &    88  \\
 AP Pup   &  30/12/00  &         500        &   119  \\
 \ldots   &  31/12/00  &         300        &   105  \\
 \ldots   &  01/01/01  &         300        &   114  \\
 \ldots   &  02/01/01  &         120        &    53  \\
 AQ Car   &  02/01/01  &         240        &    76  \\
 AQ Pup   &  01/01/01  &         500        &    84  \\
 AT Pup   &  30/12/00  &         600        &    73  \\
 AX Vel   &  31/12/00  &         400        &    94  \\
 BG Vel   &  30/12/00  &         600        &   100  \\
 BN Pup   &  02/01/01  &         600        &    62  \\
 DR Vel   &  01/01/01  &         600        &    95  \\
 lCar     &  29/12/00  &         180        &    42  \\
 \ldots   &  30/12/00  &         120        &    41  \\
 \ldots   &  31/12/00  &          80        &    40  \\
 \ldots   &  01/01/01  &         120        &    37  \\
 MY Pup   &  31/12/00  &         100        &    94  \\
 RS Pup   &  31/12/00  &         240        &    48  \\
 RY CMa   &  29/12/00  &         240        &    84  \\
 \ldots   &  30/12/00  &         600        &    67  \\
 RY Vel   &  29/12/00  &         600        &    69  \\
 RZ CMa   &  02/01/01  &         600        &    77  \\
 RZ Vel   &  01/01/01  &         300        &    42  \\
 ST Vel   &  02/01/01  &         600        &    80  \\
 SW Vel   &  31/12/00  &         500        &   153  \\
 SX Vel   &  31/12/00  &         400        &    82  \\
 T Vel    &  31/12/00  &         400        &   106  \\
 TW CMa   &  02/01/01  &         600        &    66  \\
 UX Car   &  31/12/00  &         400        &   105  \\
 V397 Car &  02/01/01  &         300        &    86  \\
 V Car    &  29/12/00  &         600        &    78  \\
 V  Vel   &  31/12/00  &         300        &   136  \\
 VX Pup   &  02/01/00  &         300        &    77  \\
 VY Car   &  31/12/00  &         240        &   130  \\
 VZ Pup   &  01/01/01  &         600        &    55  \\
 WX Pup   &  01/01/01  &         300        &    53  \\
\hline
			       
\end {tabular}					       
\label {obslog}					       
\end {table*}					       

%
   
\section{Methodology}

The determination of an accurate temperature is a critical point in the abundance 
determination. In this investigation, the temperature was derived 
spectroscopically and does not rely on photometric indices. 

\subsection{Line list}

For iron, we used the line list proposed by \citet{Mot07}, which
contains 246 Fe I lines and 17 Fe II lines covering a wide spectral 
range corresponding to the FEROS one. These lines are a compilation 
of the line lists from \citet{Cle95}, \citet{FC97}, \citet{Kiss00} and \citet{And02a}.
They were completed with a selection of lines from VALD 
(Vienne Atomic Line Database) \citep{Kup99,Rya99}, picked for effective 
temperatures corresponding to Cepheids: 4500-6500 K.

Mottini et al. devised the following approach to removing blended lines.
   \begin{itemize}
      \item All the lines stored in VALD between 4800 and 7900 \AA~and 
corresponding to Cepheids typical parameters have been compiled.
      \item This compilation was over-plotted onto the iron lines, with special 
attention given to the lines located within $\pm$ 3 \AA~of an 
iron line.   
      \item If the these lines contributed to more than 5 \%  of the iron 
line strength, then the iron line was neglected. \end{itemize}
This test was performed on three observed spectra for Cepheids with effective 
temperatures about 4500 K, 5500 K, and 6500 K. It might remain a few slightly 
blended lines, since a few weak lines may not be listed in VALD, but their 
overall contribution is negligible.

The same method  was adopted to clean the line list of alpha elements,
which are much less numerous: 4 lines for $AlI$, 9 lines for $MgI$, 4
lines for $NaI$, 2 lines for $0I$, 20 lines for $CaI$ and 24 lines 
for $SiI$. For all the lines, the physical properties (oscillator 
strength, excitation potential) listed in VALD were adopted. 

\subsection{Equivalent widths}

The measurement of equivalent widths (EW) was performed with a
semi-interactive code ($fitline$) developed by P. Fran\c cois.
This code is based on genetic algorithms from \citet{Char95}, and
it mimics the way genetic mutations affect DNA, driving the 
evolution of species. Lines are fitted by a Gaussian defined 
by four parameters: central wavelength, width and depth of the line, 
and the continuum value. Line after line, the algorithm runs in
the following way:
   \begin{enumerate}
      \item An initial set of Gaussians is computed, giving random
      values to the four parameters.
      \item The quality of each fit is estimated by calculating the $\chi^{2}$.   
      \item A new "generation" of Gaussians is calculated from the
      20 best fits after adding random modifications to the initial set of 
      parameters values (``mutations''). 
      \item The new set replaces the old one, and its accuracy is estimated 
by once again using a $\chi^{2}$ evaluation.
      \item The process is iterated 100 to 200 times until the convergence to the 
best Gaussian fit is attained.
   \end{enumerate}
In the current investigation we limit our analysis to weak, non blended lines, all 
of which could be fitted by a Gaussian. The strong lines were neglected. 
Some weak but asymmetric lines were also neglected.  This selection reduced 
the number of lines really used to 30-90 for $FeI$, 5-10 for $FeII$, 1-3 for 
$O$, $Na$, $Mg$ and $Al$, 1-10 for $Ca$, and 10-20 for $Si$. 

\subsection{Temperature determination}

We obtained the stellar effective temperature T$_{eff}$ by using
the method of line depth ratios, described in \citet{KovGor00}.
This method only relies on spectroscopy and, therefore, is independent 
of interstellar reddening. The key advantage to this approach is that 
it uses weak, neutral metallic lines appropriately selected:
   \begin{itemize}
      \item The excitation potentials of line depth ratios must differ 
as much as possible for a good estimation of T$_{eff}$, but lines should be 
relatively weak to limit the dependence on microturbulent velocity. 
      \item Pairs of lines must be close enough to eliminate errors in continuum placement. 
      \item Ionized lines are excluded because they are too sensitive to the surface gravity.  
      \item The line depth ratios are based on weak lines, and thus metallicity effects on 
temperature determination are strongly moderated. It is worth noting that two lines located 
on different parts of the curve of growth do not have the same behaviour when metallicity 
changes.
   \end{itemize}
\citet{KovGor00} proposed 32 analytical relations for determining T$_{eff}$ from the line 
depth ratios. After measuring the line depths and calculating subsequent ratios, we 
obtained 32 temperatures whose mean value gave us the stellar temperatures listed in 
Table~\ref{atmpara}. According to the quoted authors, the uncertainties on absolute 
temperatures are of the order of 80 K, and uncertainties on relative temperatures 
inside a sample are smaller and of the order of 10-15 K. 

\subsection{Atmospheric parameters}

\begin{figure*}[ht]
	\includegraphics[angle=-90]{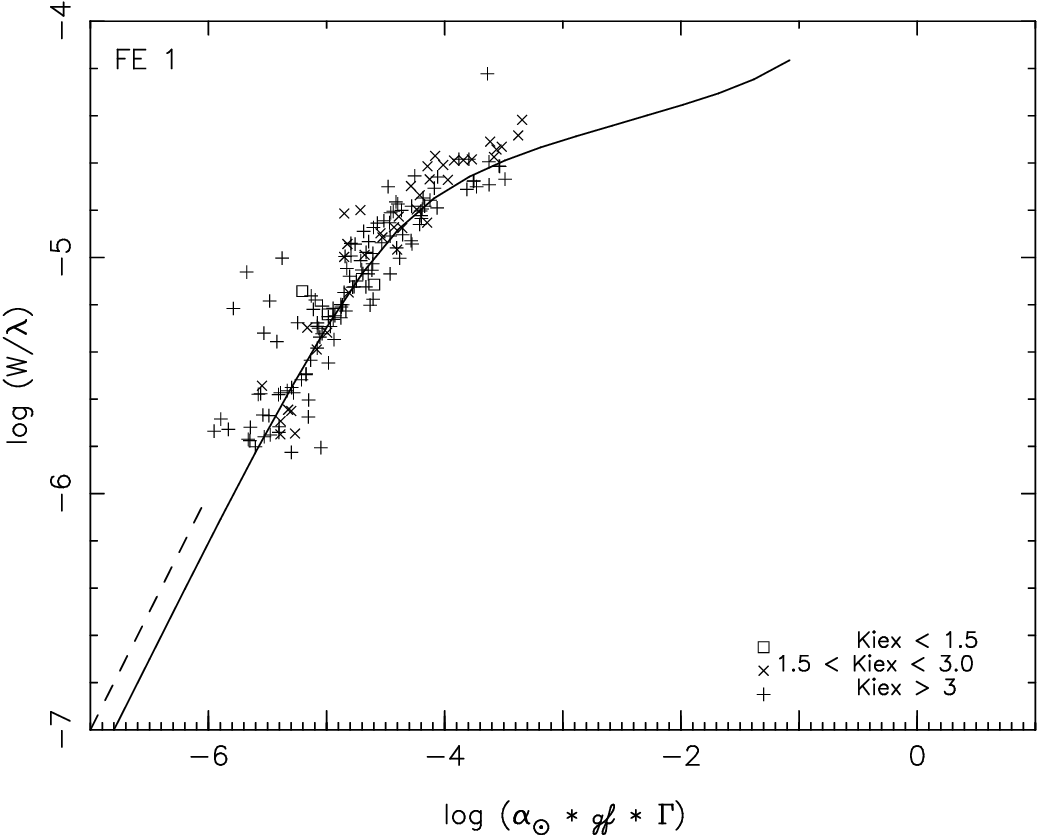}
	\caption{ Observed curve of growth for Fe I in APPup. The full line is the theoretical curve of growth for a typical line ($\lambda$=5000\AA, $\chi_{ex}$=3).
The atmospheric parameters adopted for this star are T$_{eff}$=6180 K, log g=2.2, v$_{t}$=4.0 km/s, and [Fe/H]=-0.2 dex. The dashed line on the left shows the [Fe/H]=0 location. }
        \label{APPup_curvegrowth_FE1}
\end{figure*}

\begin{figure*}[!ht]
	\includegraphics[width=14cm,angle=-90]{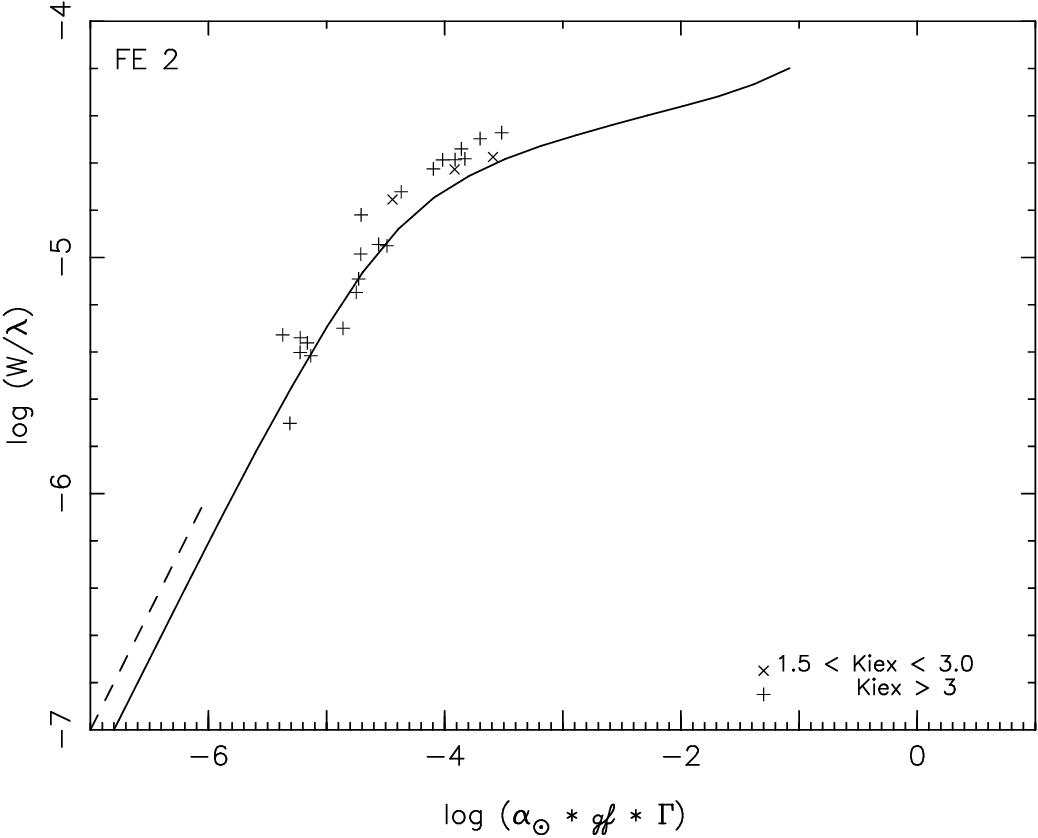}
	\caption{ Observed curve of growth for Fe II in APPup. The full line is the theoretical curve of growth for a typical line ($\lambda$=5000\AA, $\chi_{ex}$=3).
The atmospheric parameters adopted for this star are T$_{eff}$=6180 K, log g=2.2, v$_{t}$=4.0 km/s, and [Fe/H]=-0.2 dex. The dashed line on the left shows the [Fe/H]=0 location. }
        \label{APPup_curvegrowth_FE2}
\end{figure*}

The surface gravity, {\bf$log~g$}, and the microturbulent velocity v$_{t}$ are
determined by imposing the ionization balance between Fe I and Fe II with 
the help of the curves of growth. Lines of the same element in different 
ionization states should give the same abundance value. Iterations on the 
values of $log g$ and v$_{t}$ are repeated until Fe I and Fe II adjust to 
the same curve of growth. Figures~\ref{APPup_curvegrowth_FE1} 
and~\ref{APPup_curvegrowth_FE2} show the curves of growth for Fe I and Fe II 
in AP~Pup. An independent support to the T$_{eff}$ value is given by the 
evidence that both high and low $\chi_{ex}$ lines properly fit the curve 
of growth. Once abundances were calculated, atmospheric parameters were 
checked and in particular, the Fe I abundances depend neither 
on line strength nor on excitation potential. The atmospheric parameters 
for the target Cepheids are listed in Table~\ref{atmpara}.

\subsection{Abundance determination}

Abundances are then calculated with the help of the atmospheric model of 
\citet{Spi67} based on the grid of models of \citet{Gus75}. These models 
are based on the following assumptions: plane-parallel stratification, 
hydrostatic equilibrium, and LTE.
The abundance determination codes adjust abundances until matching the observed 
equivalent width. We adopted the scaled-solar chemical abundances suggested by 
\citet{Gre96}.

For every star in the sample, the final abundance was obtained by calculating 
the mean value of the abundances found for each line. The use of a substantial 
number of ``clean'' lines drives to very small error bars, which are less 
than 0.10 dex for most of them. Oxygen and aluminum abundances were checked with 
synthetic spectra, to account for the occurrence of blended lines. 
In most cases, the influence of the blended lines was negligible. 
In a few cases, abundances were modified, but the change never exceeded 
0.1 dex.

\subsection{Uncertainties in abundance determinations}

The first source of uncertainty comes from the extraction of data from the 
spectra: the equivalent width and the continuum placement (which influences 
the measure of the equivalent width) must be precisely determined. To minimize 
the errors, we limited our analysis to symmetric and unblended lines and 
neglected the most crowded parts of the spectra in selecting the line list. 
We only adopted weak or relatively weak lines (EW $\leq$ 120 m\AA) and 
assumed they could be fitted by Gaussians.

The second crucial point is to have precise temperature determinations, 
because this  intrinsic parameter strongly affects the line strength.
We used, as mentioned above, the calibration relations from \citet{KovGor00}. 
According to these authors, the uncertainties on absolute temperatures are 
at most $\approx 80$ K. Since current spectra have been secured with a 
spectral resolution of only 48000, the intrinsic dispersion in temperature 
values is on average $\approx 80-130$ K.

In order to constrain how the uncertainties affect abundance determinations, 
we finally computed abundances by adopting over or under-estimated values of the 
atmospheric parameters. The test was carried out with AP~Pup, one of our spectra 
with the lowest S/N (S/N=53). As expected, only errors on temperature 
had a noticeable effect: an error of 100 K on the effective temperature implies 
a difference of about 0.1 dex on abundances values, while errors on surface 
gravity of $\Delta log g= \pm0.3$ and on turbultent velocity of 
$\Delta v_{t}\pm 0.5$ km/sec cause much smaller differences, namely 0.02 
and 0.05 dex.

Finally, the LTE assumption in the atmosphere models might not be appropriate 
for supergiant stars, hence the adopted stellar atmosphere models could 
provide abundance determinations affected by deceptive systematic errors. 
This effect should be minimized, however, by the fact that we only adopted 
weak or relatively weak lines. Moreover, \citet{FC97} show that a 
canonical spectroscopic approach, using classical LTE atmosphere models, 
gives reliable abundances for Cepheids. In particular, they noted that 
LTE analysis of dwarfs and Cepheids located in the same open clusters give, 
within the errors, the same metallicity.  \citet{Yong06} used the same method 
(classical LTE models) and examined the difference [TiI/Fe] - [TiII/Fe] to 
test the influence of NLTE  effects. The discrepancy between the two different 
approaches is -0.07 dex and lies within the measurement uncertainties, thus 
suggesting that the surface gravities derived from ionization equilibrium of 
Fe spectra are robust.

\begin {table*}[!ht]
\scriptsize 
\caption {Atmospheric parameters of the Cepheid sample
with periods from Fernie's database \citep{Fer95}.} 

\begin {tabular}{lcccccc} 
\hline
\hline
  Object  &     P     &  phase &  T$_{eff}$   & log g & V$_{t}$ & [Fe/H] \\
          &     d     &        &      K       &       &   km/s  &   dex  \\
 \hline 
 AD Pup   & 13.594    & 0.406  & 5290$\pm$146 & 1.0 & 2.4 & -0.4  \\
 AH Vel   &  4.227171 & 0.408  & 6080$\pm$ 80 & 1.5 & 3.5 & -0.1  \\
 \ldots   &  4.227171 & 0.887  & 6483$\pm$168 & 1.6 & 3.2 & -0.1  \\
 AP Pup   &  5.084274 & 0.733  & 6180$\pm$ 84 & 2.2 & 4.0 & -0.2  \\
 \ldots   &  5.084274 & 0.926  & 6148$\pm$ 93 & 1.7 & 3.5 & -0.2  \\
 \ldots   &  5.084274 & 0.130  & 5723$\pm$ 87 & 1.6 & 3.7 & -0.2  \\
 \ldots   &  5.084274 & 0.329  & 5557$\pm$ 86 & 1.7 & 3.5 & -0.2  \\
 AQ Car   &  9.76896  & 0.528  & 5374$\pm$104 & 0.5 & 3.5 & -0.1  \\
 AQ Pup   & 30.104    & 0.309  & 5304$\pm$133 & 0.3 & 3.3 & -0.4  \\
 AT Pup   &  6.664879 & 0.594  & 5348$\pm$144 & 0.7 & 3.2 & -0.4  \\
 AX Vel   &  3.6731   & 0.598  & 5995$\pm$161 & 1.8 & 3.2 & -0.3  \\
 BG Vel   &  6.923655 & 0.895  & 5957$\pm$111 & 1.5 & 3.5 & -0.2  \\
 BN Pup   & 13.6731   & 0.116  & 5615$\pm$ 81 & 1.0 & 3.0 & -0.1  \\
 DR Vel   & 11.1993   & 0.179  & 5183$\pm$ 87 & 0.5 & 3.0 & -0.2  \\
 lCar     & 35.551341 & 0.234  & 4885$\pm$ 87 & 0.7 & 3.5 &  0.1  \\
 \ldots   & 35.551341 & 0.262  & 4848$\pm$ 91 & 0.5 & 3.5 &  0.1  \\
 \ldots   & 35.551341 & 0.290  & 4900$\pm$105 & 0.5 & 3.5 &  0.1  \\
 \ldots   & 35.551341 & 0.318  & 4821$\pm$ 87 & 0.6 & 3.2 &  0.1  \\
 MY Pup   &  5.695309 & 0.904  & 6114$\pm$110 & 1.4 & 2.8 & -0.25 \\
 RS Pup   & 41.3876   & 0.826  & 5056$\pm$110 & 0.2 & 3.2 & -0.1  \\
 RY CMa   &  4.67825  & 0.257  & 5666$\pm$131 & 1.2 & 3.0 & -0.2  \\
 \ldots   &  4.67825  & 0.473  & 5460$\pm$106 & 1.2 & 3.2 & -0.1  \\
 RY Vel   & 28.1357   & 0.424  & 5124$\pm$149 & 0.3 & 3.5 & -0.3  \\
 RZ CMa   &  4.254832 & 0.084  & 5800$\pm$186 & 1.0 & 2.7 & -0.3  \\
 RZ Vel   & 20.39824  & 0.606  & 4992$\pm$131 & 0.5 & 3.2 & -0.1  \\
 ST Vel   &  5.858425 & 0.224  & 5474$\pm$109 & 1.1 & 3.2 & -0.1  \\
 SW Vel   & 23.441    & 0.583  & 5150$\pm$ 96 & 0.5 & 5.0 & -0.2  \\
 SX Vel   &  9.54993  & 0.983  & 6248$\pm$132 & 1.3 & 2.8 & -0.2  \\
 T Vel    &  4.639819 & 0.135  & 5800$\pm$ 91 & 1.8 & 3.0 & -0.1  \\
 TW CMa   &  6.99507  & 0.471  & 5364$\pm$184 & 1.0 & 4.0 & -0.5  \\
 UX Car   &  3.682246 & 0.577  & 6002$\pm$ 97 & 1.7 & 3.5 & -0.1  \\
 V397 Car &  2.0635   & 0.765  & 5959$\pm$161 & 1.5 & 3.5 & -0.15 \\
 V Car    &  6.696672 & 0.357  & 5423$\pm$112 & 1.0 & 3.0 & -0.1  \\
 V  Vel   &  4.371043 & 0.279  & 5608$\pm$121 & 1.3 & 3.0 & -0.3  \\
 VX Pup   &  3.01087  & 0.733  & 5870$\pm$ 97 & 1.8 & 3.0 & -0.15 \\
 VY Car   & 18.913762 & 0.334  & 6127$\pm$147 & 1.5 & 4.2 & -0.05 \\
 VZ Pup   & 23.171    & 0.653  & 5015$\pm$114 & 0.2 & 4.0 & -0.3  \\
 WX Pup   &  8.93705  & 0.497  & 5263$\pm$132 & 0.7 & 3.0 & -0.1  \\
\hline								    
			       					    
\end {tabular}					       		    
\label {atmpara}					       	    
\end {table*}							    
						    
\subsection{Distances}

Distance calculations are based on NIR ($J,H,K-$band) photometry from 
\citet{LS94}. For a few objects we adopted NIR photometry from the 2MASS catalogue. 
The mean magnitude of these objects was estimated using the template light curves 
provided by \citet{Sos05}, together with the $V-$band amplitude and the 
epoch of maximum available in the literature. \citet{LS94} magnitudes have already 
been un-reddened, while for un-reddening the new data, we first computed the relative 
absorption in the $V-$band, A$_{V}$ = 3.1 E(B-V), and then according to \citet{Car89} 
the absorption in the $J$, $H$, and $K-$band, namely A$_{J}$ = 0.28 A$_{V}$, 
A$_{H}$ = 0.19 A$_{V}$, and A$_{K}$ = 0.114 A$_{V}$. For individual Cepheids we 
adopted the reddening values given in the Fernie's database \citep{Fer95}.

The Cepheid absolute magnitudes were computed using the NIR period-luminosity 
(PL) relations recently provided by \citet{Per04}, together with an LMC distance 
modulus of 18.50. Pulsation periods are from the Fernie's database \citep{Fer95}. 
The period of Cepheids pulsating in the first overtone were fundamentalized,
while we adopted the fundamental period for the double-mode pulsators 
pulsating simultaneously in the fundamental and in the first overtone.
The PL relations provided by \citet{Per04} are based on a large and homogeneous 
sample of LMC Cepheids, but it can be applied to Galactic Cepheids. Indeed, at 
NIR wavelengths, the slopes and the zero-points of the PL relation depend very 
weakly on metallicity \citep{Bono99,Per04,Mot07}. Moreover, the NIR PL relations 
present an intrinsic dispersion that is significantly smaller 
than the dispersion of the optical PL relations. The difference is due to the 
fact that the former ones are, at fixed period,  marginally affected by the 
intrinsic width in temperature of the instability strip. Cepheid distances 
based on NIR PL relations are more accurate, when compared with optical ones, 
because they are also marginally affected by the uncertainties in reddening 
estimates.   

The heliocentric distance $d$ can now be computed:
\begin{center}
d = 10$^{0.2 (m^{0}_{X}-M_{X}+5)}$ , where x = J, H, K.
\end{center}
The heliocentric distance adopted in the following is the mean value between 
the three distances in $J$, $H$, and $K-$ bands to limit random errors. 
The Galactocentric distance is finally calculated with the classical formula:
\begin{center}
R$_{G}$ = [R$_{G,hel}^{2}$ + (d cosb)$^2$ - 2 R$_{G,hel}$ d cos b cos l ]$^{1/2}$
\end{center}
where R$_{G,hel}$ is the Galactocentric distance of the Sun, $d$ the
heliocentric distance of the Cepheid, while $l$ and $b$ are the Galactic 
longitude and latitude, listed in the Fernie database \citep{Fer95}. 
We assumed a galactocentric distance of 8.5 kpc for the sun \citep{Ker86,Fea97}.

The results relative to current Cepheid sample are given in Table \ref{dist}. 
Distances based on NIR photometry are compared with distances based on the 
catalogue of light-curve parameters and distances provided by \citet{Ber00}. 
In this catalogue, the heliocentric distance was calculated using a 
period-colour relation and photometric measurements in the optical bands. 
Results show good agreement for Cepheids near the solar circle and are 
slightly more discrepant toward the outer disk, where the differences can 
reach 0.3 kpc and even 0.6 kpc (5\%) for AD Pup, the most distant Cepheid 
in our sample.

\begin {table*}[!ht]				

\scriptsize 						
\caption {Cepheids distances from IR photometry (d$_{IR}$) and comparison with the distances computed from \citet{Ber00} (d$_{B}$).}
						
\begin {tabular}{lccc} 				
\hline						
\hline						
  Object  & d$_{IR}$ & d$_{B}$ \\
          &   kpc    &   kpc   \\        
 \hline           				
 AD Pup   &  11.64   & 11.03   \\
 AH Vel   &   8.59   &  8.60   \\
 AP Pup   &   8.81   &  8.75   \\
 AQ Car   &   8.20   &  8.18   \\
 AQ Pup   &  10.15   & 10.02   \\
 AT Pup   &   9.07   &  8.98   \\
 AX Vel   &   8.75   &  8.71   \\
 BG Vel   &   8.52   &  8.51   \\
 BN Pup   &  10.64   & 10.35   \\
 DR Vel   &   8.62   &  8.57   \\
 lCar     &   8.40   &  8.40   \\
 MY Pup   &   8.61   &  8.62   \\
 RS Pup   &   9.19   &  9.15   \\
 RY CMa   &   9.41   &  9.26   \\
 RY Vel   &   8.32   &  8.31   \\
 RZ CMa   &   9.99   &  9.60   \\
 RZ Vel   &   8.83   &  8.79   \\
 ST Vel   &   8.78   &  8.68   \\
 SW Vel   &   9.06   &  8.98   \\
 SX Vel   &   8.90   &  8.84   \\
 T Vel    &   8.66   &  8.62   \\
 TW CMa   &  10.46   & 10.22   \\
 UX Car   &   8.25   &  8.27   \\
 V397 Car &   8.27   &  8.27   \\
 V Car    &   8.48   &  8.47   \\
 V  Vel   &   8.44   &  8.44   \\
 VX Pup   &   9.38   &   -     \\
 VY Car   &   8.16   &  8.18   \\
 VZ Pup   &  11.56   & 11.35   \\
 WX Pup   &   9.98   &  9.80   \\
\hline	                    				
	                    	       			
\end {tabular}					
\label {dist}					
\end {table*}				

%
					    
\section{Cepheid metal abundances}								    

Exhaustive studies of Cepheid chemical compositions are recent after 
rapidly developing since the introduction of high-resolution, wide 
spectral-range spectrographs. Precise values of Cepheid abundances 
are necessary for constraining the metallicity dependence of optical 
and NIR PL relations and for determining Galactic abundance gradients.
This investigation deals with accurate abundances of iron and 
several light metals for a good sample of Galactic Cepheids. The 
results are listed in Tables 4 and 5, and for each target we give 
the abundances and the standard deviations (when more than 1 line 
was adopted). These tables also include previous abundance measurements 
when available. The literature abundances come mainly 
from the systematic investigation by Andrievsky and collaborators
who published accurate chemical abundances for a large sample of Galactic 
Cepheids in a series of papers: \citet{And02a,And02b,And02c}, 
\citet{Luck03}, \citet{And04}.

Current measurements, when compared with the quoted abundance analyses, 
present on average slightly better standard deviations. We typically used 
a smaller number of lines, but their fits have been interactively 
checked one-by-one. The standard deviations are on average smaller than 
0.10 dex for $Fe$ and $Na$, and they range from 0.05 to 0.20 dex for $O$, 
$Al$, $Mg$, $Si$ and $Ca$. As a whole, we found that Cepheids in the 
solar neighbourhod have solar-like abundances both for iron and 
$\alpha-$elements, as expected from previous abundance analyses on 
Cepheids and supergiants 
\citep{Luck89,Luck92,Luck95,And02a,And02b,And02c,Luck03,And04,Kov05}.

\subsection{Iron}
\begin{figure*}[!ht]
	\includegraphics[width=14cm,angle=-90]{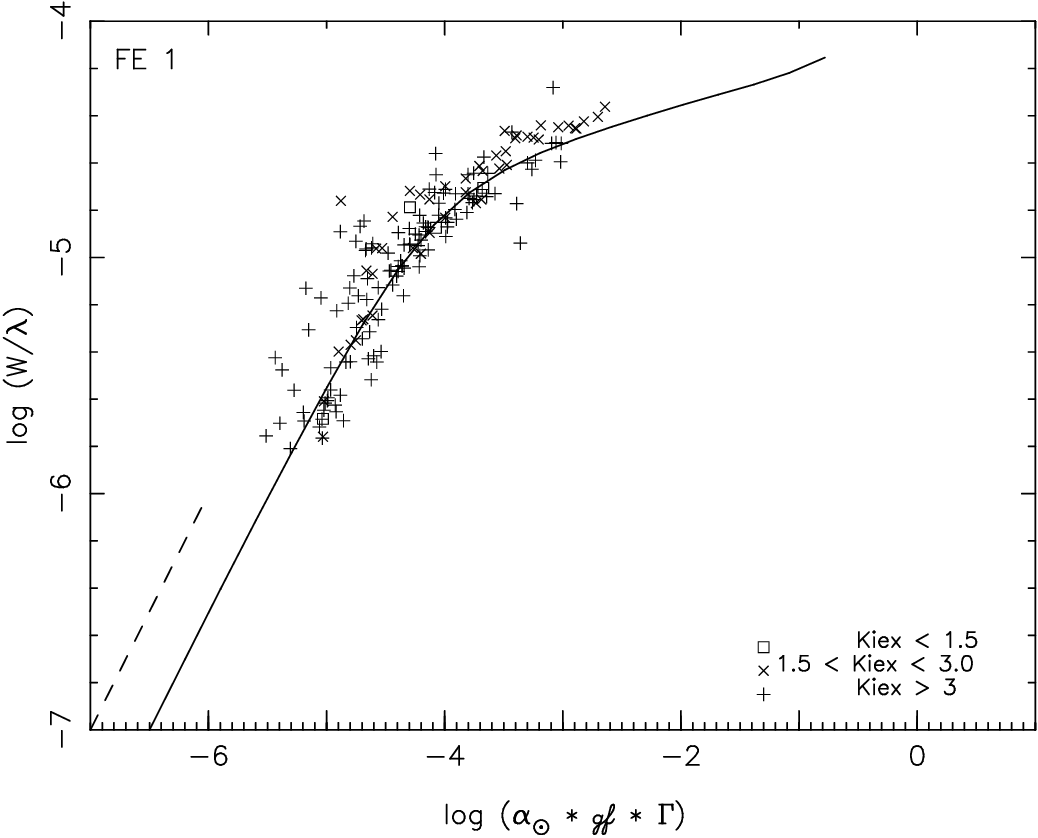}
	\caption{ Observed curve of growth for Fe I in TW~CMa. The full line represents the theoretical curve of growth for a typical line ($\lambda$=5000\AA, $\chi_{ex}$=3).
The atmospheric parameters adopted for this star are T$_{eff}$=5364 K, log g=1.0, v$_{t}$=4.0 km/s, and [Fe/H]=-0.5 dex. The dashed line on the left shows the [Fe/H]=0 location. }
        \label{TWCMa_curvegrowth_FE1}
\end{figure*}

\begin{figure*}[!ht]
	\includegraphics[width=14cm,angle=-90]{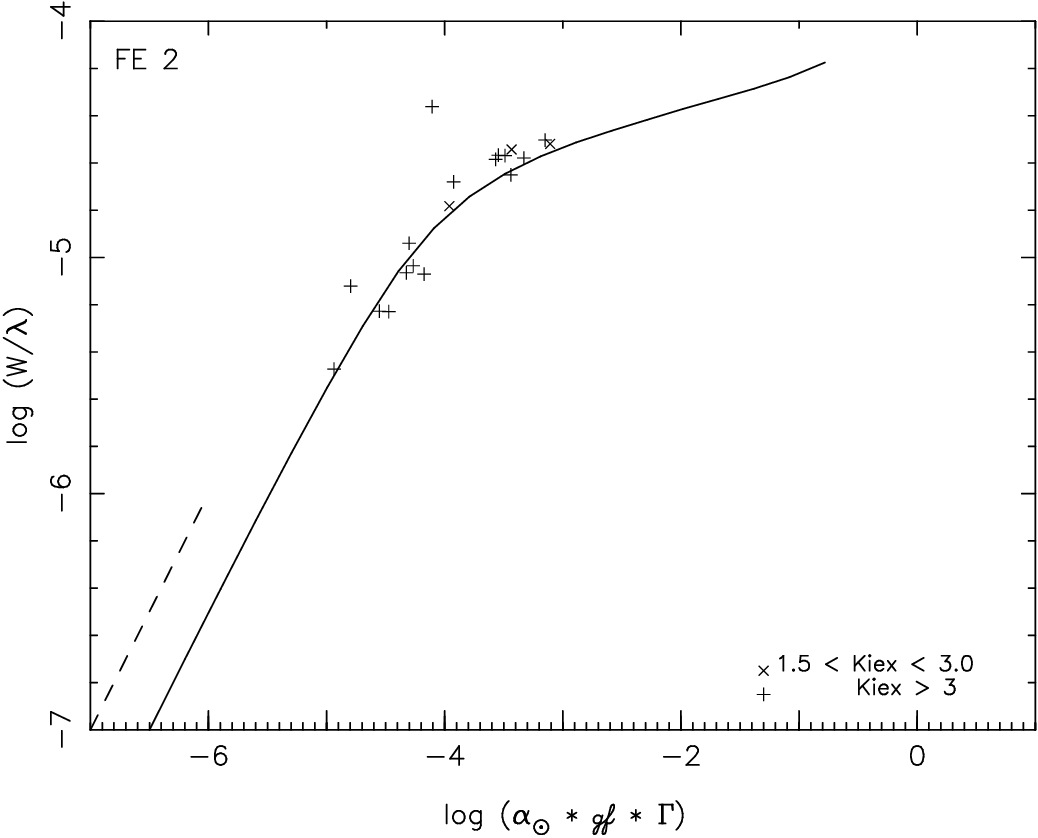}
	\caption{ Observed curve of growth for Fe II in TW~CMa. The full line represents the theoretical curve of growth for a typical line ($\lambda$=5000\AA, $\chi_{ex}$=3).
The atmospheric parameters adopted for this star are T$_{eff}$=5364 K, log g=1.0, v$_{t}$=4.0 km/s, and [Fe/H]=-0.5 dex. The dashed line on the left shows the [Fe/H]=0 location. }
        \label{TWCMa_curvegrowth_FE2}
\end{figure*}

\begin{figure*}[!ht]
	\includegraphics[width=14cm,angle=-90]{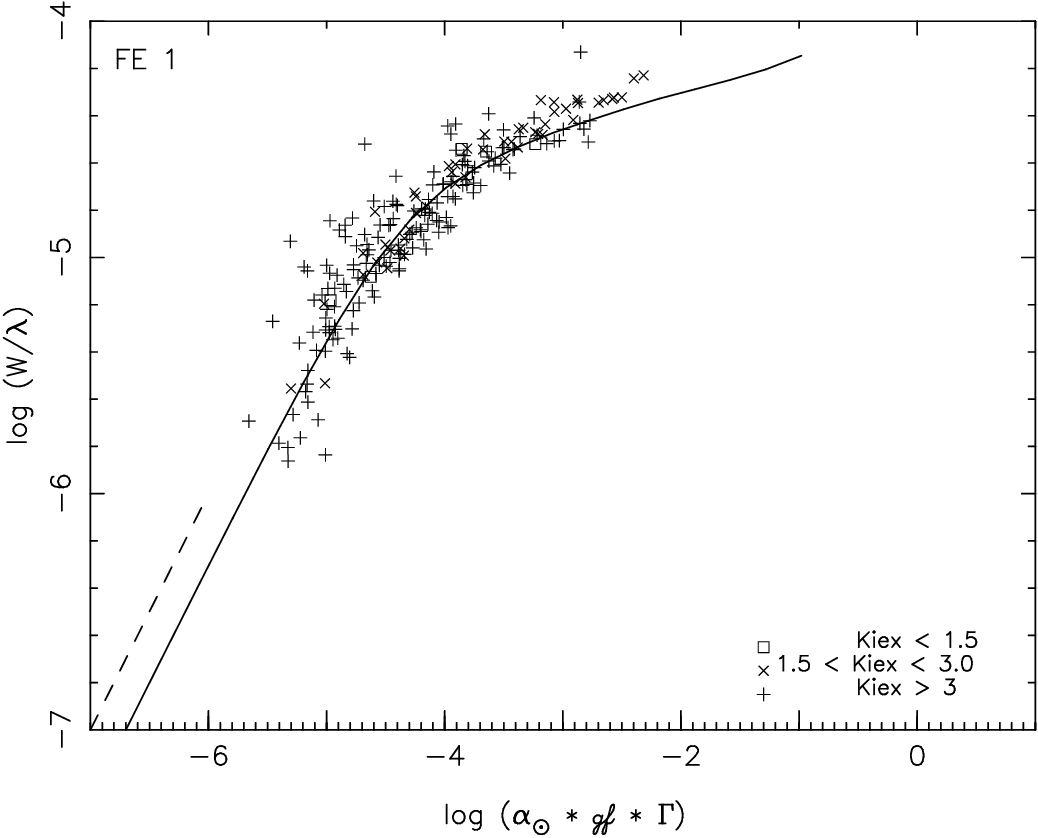}
	\caption{Observed curve of growth for Fe I in VZPup. The full line represents the theoretical curve of growth for a typical line ($\lambda$=5000\AA, $\chi_{ex}$=3);
The atmospheric parameters adopted for this star are T$_{eff}$=5015 K, log g=0.2, v$_{t}$=4.0 km/s, and [Fe/H]=-0.3 dex. The dashed line on the left shows the [Fe/H]=0 location. }
        \label{VZPup_curvegrowth_FE1}
\end{figure*}

\begin{figure*}[!ht]
	\includegraphics[width=14cm,angle=-90]{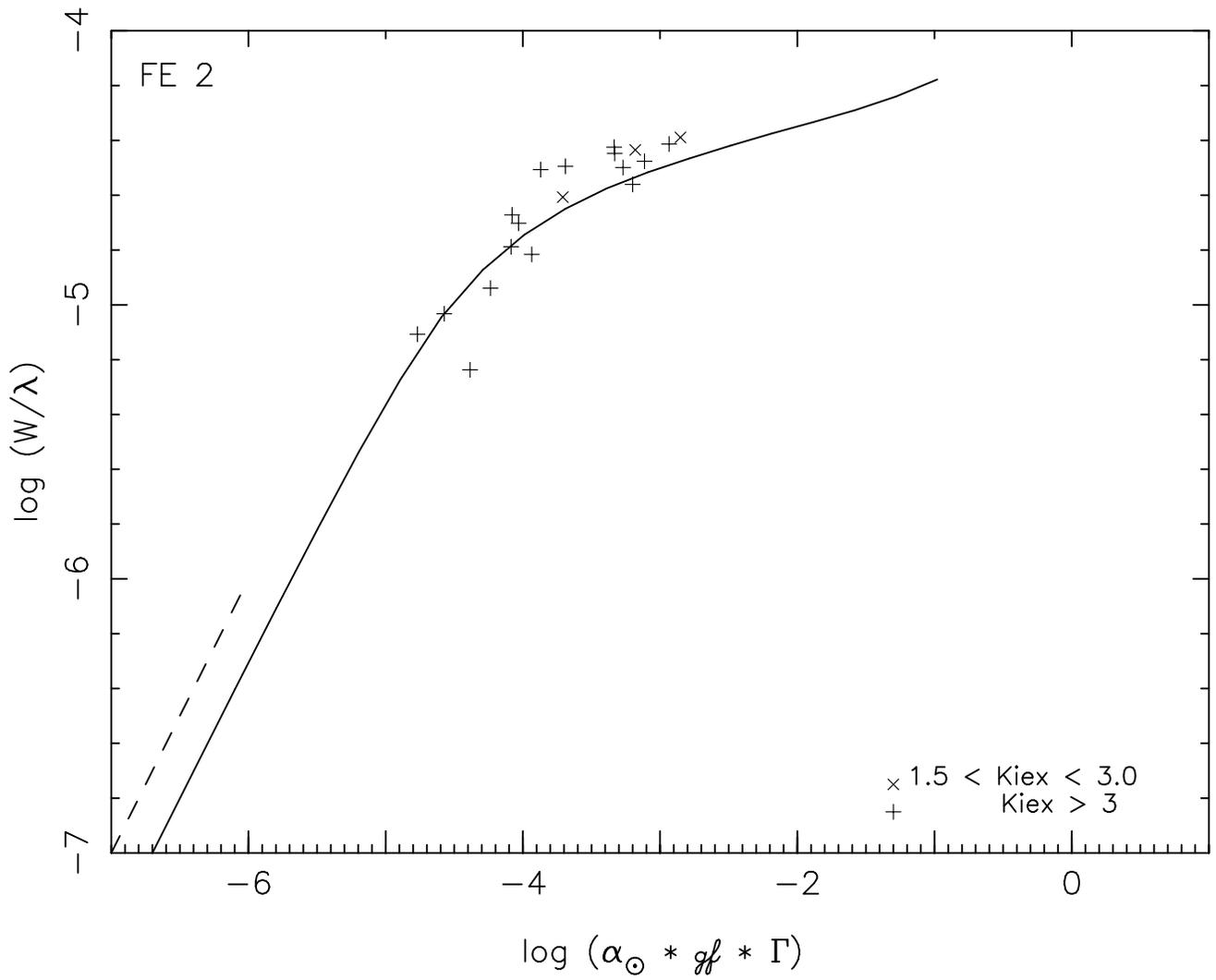}
	\caption{ Observed curve of growth for Fe II in VZPup. The full line represents the theoretical curve of growth for a typical line ($\lambda$=5000\AA, $\chi_{ex}$=3).
The atmospheric parameters adopted for this star are T$_{eff}$=5015 K, log g=0.2, v$_{t}$=4.0 km/s, and [Fe/H]=-0.3 dex. The dashed line on the left shows the [Fe/H]=0 location. }
        \label{VZPup_curvegrowth_FE2}
\end{figure*}

For iron (see Table~\ref{Fe/O/Na}), our results show excellent agreement 
with Andrievsky's, and the discrepancies are typically smaller than 0.10 dex.
Larger discrepancies ($\approx 0.2-0.3$ dex) have only been found for two 
targets, namely TW~CMa and VZ~Pup. Similar discrepancies have also been found 
for other elements. These differences might be due either to different signal-to-noise 
spectra or to the adopted atmospheric parameters. However, the curves of growth 
for these two targets and related tests appear to be very consistent.
Figures~\ref{TWCMa_curvegrowth_FE1},~\ref{TWCMa_curvegrowth_FE2} 
and~\ref{VZPup_curvegrowth_FE1},~\ref{VZPup_curvegrowth_FE2} show the 
results for TW~CMa and VZ~Pup, respectively. A glance at the data plotted in 
these figures shows good agreement between the predicted curves of growth 
and current measurements for both $FeI$ and $FeII$ lines. These distant stars 
deserve further investigations, as they are quite distant from the Galactic 
centre, and therefore very useful to firmly constrain the metallicity gradient 
in the outer zones of the Galaxy. 

\begin {table*}[!ht]

\scriptsize 
\caption {Iron, oxygen, sodium abundances, and comparison with previous abundance analyses: 
(1) \citet{And02a} and foll. (2) \citet{FC97}. (3) \citet{Luck81}. (4) \citet{Boy84}. (5) \citet{Bar82}. (6) \citet{Luck85}. 
 (7) \citet{Luck92}. (8) \citet{And94}. (9) \citet{The98} (10) \citet{Mot07}.}
\begin {tabular}{lcccccc} 
\hline
\hline
 Object   &     [Fe/H]     &           Others                      &     [O/H]      &    Others   &     [Na/H]     &    Others   \\                                   
          &      dex       &            dex                        &      dex       &     dex     &       dex      &     dex     \\
 \hline 							                                                                   
 AD Pup   & -0.20$\pm$0.12 & -0.24$^{1}$                           & -0.09$\pm$0.04 &             &  0.10$\pm$0.03 &  0.06$^{1}$ \\
 AH Vel   & -0.06$\pm$0.05 &                                       &  0.14          &             &  0.31$\pm$0.01 &             \\
 \ldots   & -0.03$\pm$0.07 &                                       & -0.01          &             &  0.25$\pm$0.05 &             \\
 AP Pup   & -0.18$\pm$0.08 & -0.07$^{10}$                          &   -            &             &  0.16$\pm$0.01 &             \\
 \ldots   & -0.13$\pm$0.07 &                                       &   -            &             &  0.05$\pm$0.04 &             \\
 \ldots   & -0.16$\pm$0.08 &                                       & -0.29          &             &  0.02$\pm$0.01 &             \\
 \ldots   & -0.15$\pm$0.08 &                                       &  0.08          &             &  0.03$\pm$0.06 &             \\
 AQ Car   & -0.30$\pm$0.10 &                                       &   -            &             &  0.16$\pm$0.01 &             \\
 AQ Pup   & -0.26$\pm$0.03 & -0.14$^{1}$; -0.09$^{10}$             & -0.27$\pm$0.04 &             &  0.08          &  0.09$^{1}$ \\
 AT Pup   & -0.22$\pm$0.10 &                                       &   -            &             &  0.21$\pm$0.04 &             \\
 AX Vel   & -0.15$\pm$0.07 & -0.67$^{5}$                           &   -            &             &  0.06$\pm$0.06 &             \\
 BG Vel   & -0.10$\pm$0.08 &                                       &   -            &             &  0.16$\pm$0.06 &             \\
 BN Pup   & -0.03$\pm$0.10 &  0.01$^{1}$; -0.09$^{10}$             & -0.01$\pm$0.01 & -0.06$^{1}$ &  0.24$\pm$0.03 &  0.09$^{1}$ \\
 DR Vel   & -0.01$\pm$0.07 &                                       & -0.41$\pm$0.09 &             &  0.25          &             \\
 lCar     &  0.11$\pm$0.07 & -0.4$^{4}$; -0.00$^{10}$              & -0.08          &             & -0.36          &             \\
 \ldots   &  0.10$\pm$0.11 & -0.3$^{6}$                            & -0.37          &             & -0.23$\pm$0.25 &             \\
 \ldots   &  0.10$\pm$0.17 &  0.02$^{7}$                           & -0.12          &             & -0.28          &             \\
 \ldots   &  0.09$\pm$0.22 & -0.16$^{9}$                           & -0.09          &             &  0.14          &             \\
 MY Pup   & -0.25$\pm$0.08 & -0.12$^{1}$                           & -0.12          & -0.12$^{1}$ &  0.10$\pm$0.01 &  0.14$^{1}$ \\
 RS Pup   &  0.07$\pm$0.09 &  0.17$^{1}$; -0.07$^{3}$; -0.12$^{10}$& -0.02$\pm$0.22 &  0.10$^{1}$ &   -            &  0.42$^{1}$ \\
 RY CMa   & -0.12$\pm$0.09 &  0.02$^{1}$                           & -0.28          & -0.30$^{1}$ &  0.09$\pm$0.07 &  0.16$^{1}$ \\
 \ldots   & -0.20$\pm$0.09 &                                       &   -            &             & -0.01$\pm$0.02 &             \\
 RY Vel   & -0.05$\pm$0.08 & -0.03$^{1}$                           & -0.22          & -0.31$^{1}$ &  0.31          &  0.20$^{1}$ \\
 RZ CMa   & -0.20$\pm$0.07 &                                       & -0.55          &             & -0.08$\pm$0.03 &             \\
 RZ Vel   &  0.05$\pm$0.10 & -0.07$^{1}$                           &   -            & -0.15$^{1}$ &  0.29          &  0.15$^{1}$ \\
 ST Vel   & -0.14$\pm$0.10 &                                       & -0.34          &             &  0.11$\pm$0.02 &             \\
 SW Vel   & -0.15$\pm$0.08 & -0.07$^{1}$; -0.08$^{2}$              &   -            &  0.18$^{1}$ &   -            &  0.17$^{1}$ \\
 SX Vel   & -0.18$\pm$0.07 & -0.03$^{1}$                           & -0.31          &  0.12$^{1}$ & -0.01          &  0.05$^{1}$ \\
 T Vel    & -0.02$\pm$0.07 & -0.02$^{1}$                           & -0.02          &  0.02$^{1}$ &  0.15$\pm$0.01 &  0.06$^{1}$ \\
 TW CMa   & -0.51$\pm$0.09 & -0.18$^{1}$                           &   -            & -0.16$^{1}$ & -0.15$\pm$0.03 & -0.03$^{1}$ \\
 UX Car   & -0.10$\pm$0.07 &                                       &   -            &             &  0.11$\pm$0.01 &             \\
 V397 Car & -0.08$\pm$0.09 &                                       &   -            &             &  0.24$\pm$0.06 &             \\
 V Car    & -0.06$\pm$0.07 & -0.04$^{10}$                          & -0.22$\pm$0.16 &             &  0.17$\pm$0.03 &             \\
 V  Vel   & -0.30$\pm$0.06 &                                       &   -            &             & -0.34$\pm$0.17 &             \\
 VX Pup   & -0.15$\pm$0.12 & -0.13$^{1}$; -0.70$^{5}$; -0.39$^{8}$ & -0.10          &             &  0.00$\pm$0.03 &  0.08$^{1}$ \\
 VY Car   & -0.06$\pm$0.06 &                                       &   -            &             &  0.15$\pm$0.21 &             \\
 VZ Pup   & -0.37$\pm$0.07 & -0.16$^{1}$                           & -0.53          & -0.10$^{1}$ & -0.08$\pm$0.01 &  0.00$^{1}$ \\
 WX Pup   & -0.15$\pm$0.09 &                                       & -0.36          &             & -0.09          &             \\
\hline								    
			       					    
\end {tabular}
\label {Fe/O/Na}					       	    
\end {table*}							   

\subsection{Oxygen}	

The oxygen is slightly deficient, as expected after the investigations by 
\citet{Luck81,Luck85}, \citet{Luck89}, and the theoretical work of \citet{Scha92}.
Current results show very good agreement with $O$ abundances provided 
by \citet{And04}, but with the exception of SX~Vel and once again VZ~Pup. 
The ratio [O/Fe] is very close to the abundances provided by \citet{Luck95} 
and \citet{Kov05}. However, oxygen abundances show a higher dispersion 
when compared with other elements because the lines we used 
($\lambda= 6300$ \AA~and 6363 \AA) are weak and blended. In order to 
reduce ther intrinsic dispersion we also adopted synthetic spectra, 
but the improvement was limited, since the wavelength region across these
lines is relatively noisy in most of our spectra.

\subsection{Sodium}

Abundances for sodium cover the range [-0.36 dex --- +0.31 dex]. Once again, 
they agree quite well with previous abundance measurements, and indeed the 
difference is 0.1 dex or less. Only two stars show a discrepancy of 
$\sim 0.15$ dex, namely BN~Pup and RZ~Vel. Most of the Cepheids in our sample 
are $Na-$overabundant, a now well-known feature for Galactic supergiants. 
This $Na-$overabundance is given by intermediate-mass MS stars, the progenitors 
of yellow supergiants, by the synthesis of sodium in the $Ne-Na$ cycle in their 
convective cores. A rotationally induced turbulent and diffusive mixing brings this 
material enriched in $Na$ into the radiative envelope and shows up at the stellar 
surface soon after the first dredge up \citep{Sas86,Den94}. 

\subsection{Aluminium and magnesium}

Aluminium and magnesium are both under-abundant in our study, as expected for Galactic 
supergiants. As far as the $Mg$ is concerned, the comparison is possible only for a 
few targets, and the abundances we obtained are quite similar to 
Andrievsky's abundances. On the other hand, the aluminum abundances present 
a larger discrepancy and current measurements are systematically 0.1--0.2 dex 
lower than those by Andrievsky and collaborators. This systematic difference only 
applies to the $Al$ abundances, and it could be due to a difference in the atomic 
parameters chosen for aluminium lines.

\begin {table*}[!ht]

\scriptsize 
\caption {Aluminium, magnesium, silicon, calcium abundances and comparisons with previous 
abundace analyses: (1) \citet{And02a}.} 

\begin {tabular}{lcccccccc} 
\hline
\hline
  Object   &     [Al/H]     &    Others   &     [Mg/H]     &    Others   &     [Si/H]     &    Others   &     [Ca/H]     &    Others   \\
           &      dex       &     dex     &      dex       &     dex     &      dex       &     dex     &      dex       &     dex     \\
 \hline 								                                                                 
 AD Pup    & -0.22$\pm$0.04 &  0.00$^{1}$ & -0.41          &             & -0.04$\pm$0.15 & -0.19$^{1}$ & -0.12$\pm$0.08 & -0.17$^{1}$ \\
 AH Vel    & -0.16$\pm$0.03 &             &   -            &             &  0.04$\pm$0.16 &             & -0.13$\pm$0.12 &             \\
 \ldots    & -0.13$\pm$0.04 &             &  0.04$\pm$0.14 &             &  0.09$\pm$0.21 &             & -0.06$\pm$0.08 &             \\
 AP Pup    & -0.27          &             &  0.01          &             & -0.05$\pm$0.12 &             & -0.12$\pm$0.06 &             \\
 \ldots    & -0.17$\pm$0.06 &             & -0.01          &             &  0.05$\pm$0.14 &             & -0.14$\pm$0.04 &             \\
 \ldots    & -0.22$\pm$0.20 &             &   -            &             &  0.03$\pm$0.14 &             & -0.13$\pm$0.05 &             \\
 \ldots    & -0.24$\pm$0.03 &             &   -            &             & -0.03$\pm$0.15 &             &   -            &             \\
 AQ Car    & -0.19$\pm$0.05 &             &   -            &             & -0.11$\pm$0.24 &             & -0.34$\pm$0.47 &             \\
 AQ Pup    & -0.29$\pm$0.15 &  0.01$^{1}$ &   -            &             & -0.06$\pm$0.15 & -0.10$^{1}$ & -0.27$\pm$0.16 & -0.07$^{1}$ \\
 AT Pup    & -0.06          &             &   -            &             &  0.03$\pm$0.12 &             & -0.15$\pm$0.01 &             \\
 AX Vel    & -0,35$\pm$0.04 &             &  0.00$\pm$0.01 &             & -0.01$\pm$0.12 &             & -0.30$\pm$0.30 &             \\
 BG Vel    & -0.19$\pm$0.02 &             & -0.34          &             & -0.04$\pm$0.11 &             & -0.16$\pm$0.09 &             \\
 BN Pup    & -0.10$\pm$0.03 &  0.04$^{1}$ &   -            &  0.21$^{1}$ &  0.10$\pm$0.18 & -0.01$^{1}$ &  0.05$\pm$0.01 &  0.04$^{1}$ \\
 DR Vel    & -0.08$\pm$0.07 &             &   -            &             &  0.04$\pm$0.08 &             & -0.07          &             \\
 l Car     & -0.15$\pm$0.05 &             &   -            &             & -0.02$\pm$0.06 &             & -0.16          &             \\
 \ldots    & -0.43$\pm$0.20 &             & -0.51          &             & -0.16$\pm$0.15 &             & -0.27          &             \\
 \ldots    & -0.15$\pm$0.06 &             &   -            &             &  0.00$\pm$0.06 &             & -0.08          &             \\
 \ldots    & -0.13$\pm$0.06 &             &   -            &             &  0.01$\pm$0.09 &             & -0.14          &             \\
 MY Pup    & -0.25$\pm$0.13 &  0.03$^{1}$ & -0.36$\pm$0.20 & -0.36$^{1}$ & -0.02$\pm$0.09 & -0.08$^{1}$ & -0.16$\pm$0.09 & -0.13$^{1}$ \\
 RS Pup    & -0.08$\pm$0.03 &  0.17$^{1}$ &   -            &             &  0.08$\pm$0.12 &  0.11$^{1}$ & -0.02          &  0.21$^{1}$ \\
 RY CMa    & -0.10$\pm$0.14 &             & -0.05          &             & -0.04$\pm$0.11 &  0.04$^{1}$ & -0.09$\pm$0.04 & -0.02$^{1}$ \\
 \ldots    & -0.28$\pm$0.04 &             & -0.26          &             & -0.11$\pm$0.13 &             & -0.28$\pm$0.17 &             \\
 RY Vel    & -0.06$\pm$0.07 &  0.12$^{1}$ &   -            & -0.26$^{1}$ &  0.06$\pm$0.12 &  0.06$^{1}$ &   -            & -0.11$^{1}$ \\
 RZ CMa    & -0.31$\pm$0.05 &             & -0.17$\pm$0.18 &             & -0.14$\pm$0.07 &             & -0.22$\pm$0.11 &             \\
 RZ Vel    & -0.04          &  0.04$^{1}$ &   -            & -0.29$^{1}$ &  0.03$\pm$0.12 &  0.07$^{1}$ & -0.13          & -0.07$^{1}$ \\
 ST Vel    & -0.23$\pm$0.15 &             & -0.14$\pm$0.00 &             & -0.01$\pm$0.11 &             & -0.56$\pm$0.49 &             \\
 SW Vel    & -0.09$\pm$0.25 &  0.04$^{1}$ &   -            &  0.16$^{1}$ &   -            & -0.03$^{1}$ & -0.27$\pm$0.21 & -0.16$^{1}$ \\
 SX Vel    & -0.16$\pm$0.14 &  0.04$^{1}$ & -0.14$\pm$0.16 & -0.20$^{1}$ &  0.02$\pm$0.19 & -0.00$^{1}$ & -0.09$\pm$0.07 & -0.07$^{1}$ \\
 T Vel     & -0.20$\pm$0.11 &  0.08$^{1}$ & -0.26$\pm$0.00 & -0.24$^{1}$ &  0.03$\pm$0.1  & -0.00$^{1}$ & -0.06$\pm$0.17 & -0.02$^{1}$ \\
 TW CMa    & -0.47$\pm$0.13 &             & -0.45$\pm$0.00 &             & -0.30$\pm$0.14 & -0.10$^{1}$ & -0.24          & -0.19$^{1}$ \\
 UX Car    & -0.30$\pm$0.01 &             & -0.11$\pm$0.16 &             &  0.01$\pm$0.18 &             &  0.00$\pm$0.01 &             \\
 V397 Car  & -0.23$\pm$0.08 &             & -0.20          &             &  0.00$\pm$0.28 &             & -0.15$\pm$0.13 &             \\
 V Car     & -0.24$\pm$0.06 &             & -0.28$\pm$0.10 &             &  0.01$\pm$0.25 &             & -0.07$\pm$0.13 &             \\
 V  Vel    & -0.54$\pm$0.12 &             & -0.65$\pm$0.30 &             & -0.27$\pm$0.06 &             & -0.40$\pm$0.07 &             \\
 VX Pup    & -0.33$\pm$0.04 &             & -0.18$\pm$0.17 &             & -0.10$\pm$0.16 & -0.06$^{1}$ & -0.16$\pm$0.06 & -0.33$^{1}$ \\
 VY Car    & -0.16          &             & -0.23$\pm$0.OO &             &  0.00$\pm$0.24 &             &  0.03          &             \\
 VZ Pup    & -0.42$\pm$0.15 & -0.15$^{1}$ &   -            & -0.15$^{1}$ & -0.18$\pm$0.16 & -0.09$^{1}$ & -0.35          & -0.12$^{1}$ \\
 WX Pup    & -0.29$\pm$0.04 &             &   -            &             & -0.04$\pm$0.18 &             & -0.21          &             \\
\hline		        					    
			       					    
\end {tabular}					       		    
\label {Al/Mg/Si/Ca}					       	    
\end {table*}

\subsection{Silicon and calcium}

These elements have, as expected, solar-like abundances. Current results 
agree quite well with the literature values, and indeed the discrepancies are 
smaller than 0.1 dex, when compared with the measurements provided by 
Andrievsky and collaborators. In a few cases, the difference is larger, 
reaching $\sim 0.2$ dex for $Si$ abundances in TW~CMa and for $Ca$ 
abundances in AQ~Pup, RS~Pup, and VX~Pup.

%

\section{Galactic radial abundance gradients}

\citet{Twa97} suggests that there is a discontinuity in the 
Galactic chemical abundance gradient in the region located at 10-11 kpc from 
the Galactic centre. On the other hand, \citet{And04} decided to adopt a 
multi-modal gradient: they divided the range they studied into three regions, 
namely  Zone I: 4.0-6.6 kpc, zone II: 6.6-10.6 kpc, and zone III: 10.6-14.6 kpc. 
Zone I shows a very steep gradient (-0.128 dex kpc$^{-1}$), the gradient 
becomes flatter in Zone II (-0.044 dex. kpc$^{-1}$), while Zone III 
shows marginal evidence of a radial gradient (+0.004 dex kpc$^{-1}$).
According to these authors, there is a jump in the metallicity gradient in 
the 10-11 kpc region, where the iron abundance shows a rapid decrease of -0.2 dex.
More recent studies based on open clusters support this working hypothesis. 
In particular, \citet{Yong05} find that, instead of steadily decreasing with 
Galactocentric distance, the metallicity reaches a basement metallicity 
($[Fe/H]\approx -0.5$ dex) beyond 10-12 kpc. This feature is also confirmed 
by the studies of 3 field red giants \citep{Carn05} and of Cepheids 
\citep{Yong06} in the outer disk. However, in the last paper, a sub-sample 
of Cepheids reaches the basement metallicity at $[Fe/H]\approx 0.9$ dex.
By using open clusters as chemical composition tracers, \citet{Twa06} 
found a bimodal gradient: the $[Fe/H]$ is close to zero for R$_G$ $<$ 10 kpc, 
and approximately -0.3 dex for larger Galactocentric distances and with a 
marginal evidence of chemical composition gradients inside the two zones.

The current Cepheid sample includes only a few targets in the crucial 
zone, and the use of the abundances estimated by \citet{And02a} and by 
\citet{Mot07} does not allow us to provide sound constraints on the 
occurrence of a discontinuity in the abundance gradients. Therefore, 
we decided to adopt a conservative approach by only computing linear 
gradients. Abundance gradients have then been calculated with a linear 
regression between abundances and  Galactocentric distances for 5 of 
the 7 elements analysed in this investigation. To compute the goodness-of-fit, 
we did not adopt the errors listed in Tables \ref{Fe/O/Na} and \ref{Al/Mg/Si/Ca} 
because they only represent the RMS of the abundances when we adopt more than 
one line and not the real measurement errors that come from uncertainties 
in the determination of the atmospheric parameters.\\
The measurement errors have been fixed as follow:
\begin{itemize}
	\item In the case of iron, we assigned them the value of 0.10 dex, 
since iron abundances have been determined with a sufficient number of lines, 
and the RMS estimates closely match the measurement errors.
This value of 0.10 dex, greater than the RMS for most of our stars, is 
therefore slightly overestimated.    
	\item For the other elements, for which the number of lines used ranges 
from 1 and 5 for a large fraction of the targets, we assigned the value of 
0.15 dex to the measurement errors. Such a value is very appropriate for $O$ 
and $Ca$, and slightly overestimated for $Na$.  \end{itemize}

To improve the sampling across the Galactic disk, we also included 
58 Cepheids from the Andrievsky's sample and 11 Cepheids from \citet{Mot07}
for which NIR photometry is available in the literature. For the former 
sample both iron and heavy element abundances are available, while for 
the latter only the iron abundances. The individual distances were 
estimated using the same approach adopted in Sect. 3.7 and the 
Galactocentric distances of these additional Cepheids are listed 
in Table \ref{adddist}. Seven of them are quite distant Cepheids 
located between 10 and 11.4 kpc, and a few are located toward the 
Galactic centre, thus extending the baseline we adopt to estimate 
the gradients. The measurement errors for these objects have been 
estimated by using the same assumptions we adopted for our sample.

\subsection{Iron Galactic radial gradient}

\begin{figure*} [!htb]
\begin{center}
\includegraphics[angle=90,width=20cm,angle=-90] {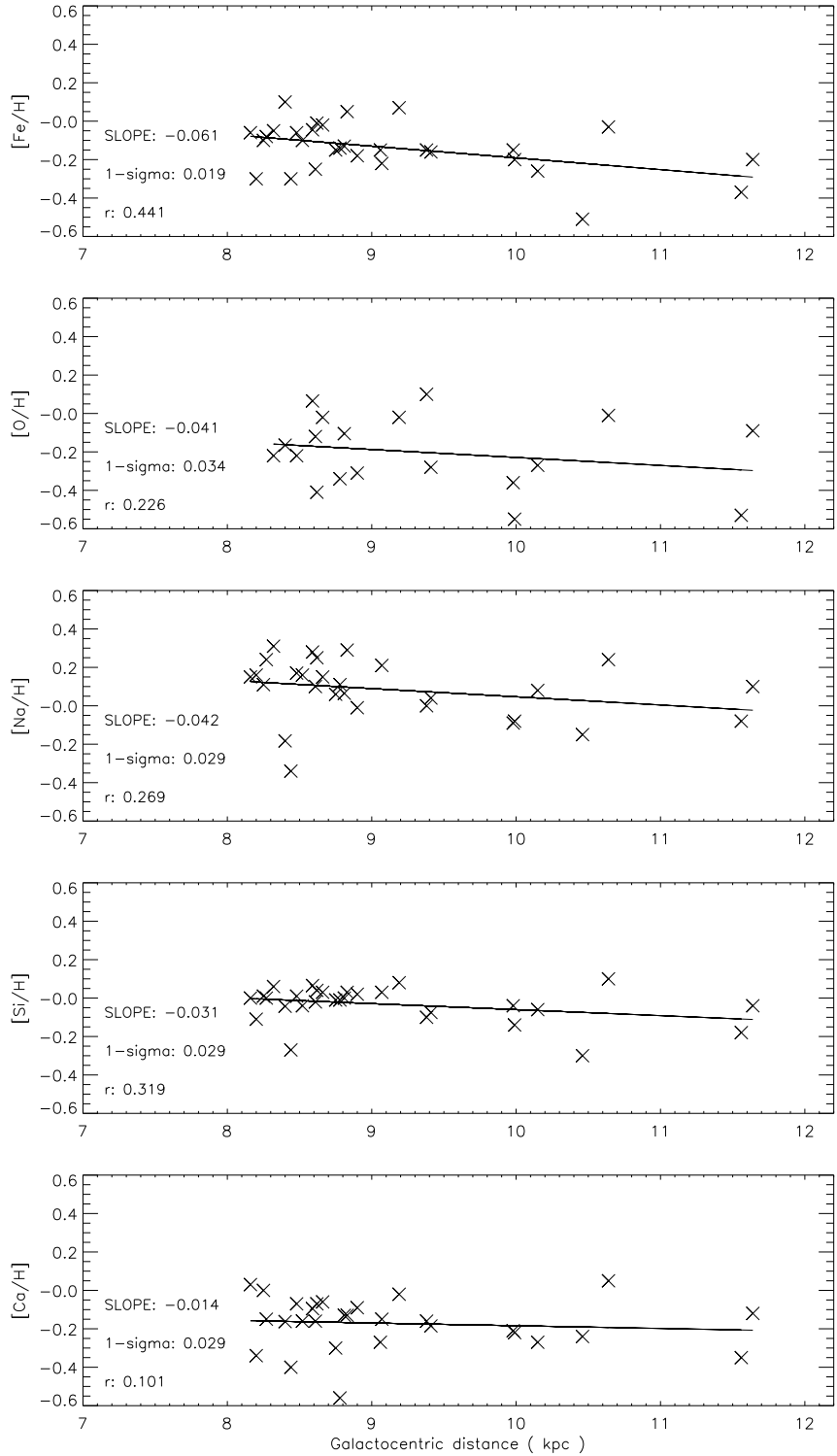}
\caption{Galactic radial abundance gradients in the [8-12] kpc region based on our 
Cepheid sample. The slopes, the $1-\sigma$ errors, and the correlation coefficients 
are also labelled. The solid line shows the linear regression.}
\label{fig:our_data}
\end{center}
\end{figure*}

We estimated at first the linear Galactic radial gradients for the current 
Cepheid sample. The slopes, their $1-\sigma$ errors, and the correlation 
coefficients are listed in columns 2,3, and 4 of Table~\ref{grad}. Data plotted 
in the top panel of Fig. \ref{fig:our_data}   
show that the slope is -0.061 dex kpc$^{-1}$ in the region ranging from 
$\approx 8$ to $\approx 12$ kpc. This seems to be a relatively high value when compared 
with the slopes determined in a similar radial interval by \citet{And02a}, who 
found -0.029 dex kpc$^{-1}$ under the hypothesis of a single-zone gradient, and 
by \citet{And04}: -0.044 dex. kpc$^{-1}$ for a multimodal gradient.
We note in passing that the correlation coefficient is relatively small (r=0.44), 
but quite similar to the correlation coefficient found by \citet{And02a} (r=0.47).

\begin{figure*} [!htb]
\begin{center}
\includegraphics[angle=90,width=20cm,angle=-90] {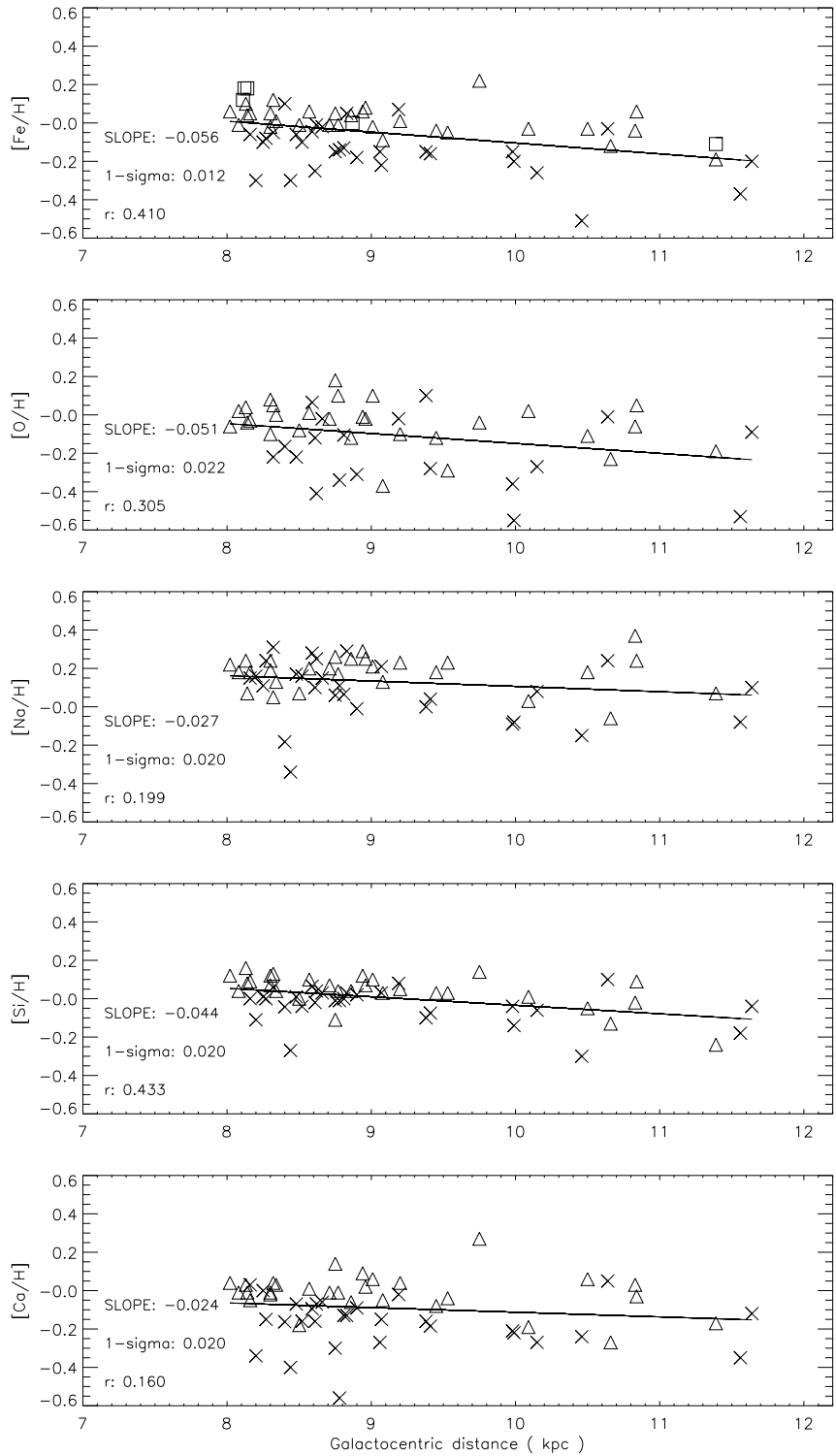}
\caption{Galactic radial abundance gradients in the [8-12] kpc region based on our 
Cepheid sample (crosses) and on data from Andrievsky and collaborators (triangles), 
and from \citet{Mot07} (squares).}
\label{fig:all_data}
\end{center}
\end{figure*}

In order to ascertain on a quantitative basis whether the current slope evaluation  
is affected by the size of the sample, we also included the Cepheids from 
Andrievsky's sample (29 Cepheids) and from the \citet{Mot07} sample 
(5 Cepheids) with Galactocentric distances ranging from $\sim 8$ 
to $\sim 12$ Kpc. This means that we doubled the sample size. 
The results of the linear regression are listed in columns 5, 6, and 7 
of Table~\ref{grad}. Interestingly enough, data plotted in the top panel of 
Fig. \ref{fig:all_data} show that the new slope, within the $1-\sigma$ error, 
agrees quite well (-0.056 vs -0.061 dex kpc$^{-1}$) with the slope only based on 
the current Cepheid sample. The same outcome applies to the correlation coefficient.  
   
\begin{figure*} [!htb]
\begin{center}
\includegraphics[angle=90,width=20cm,angle=-90] {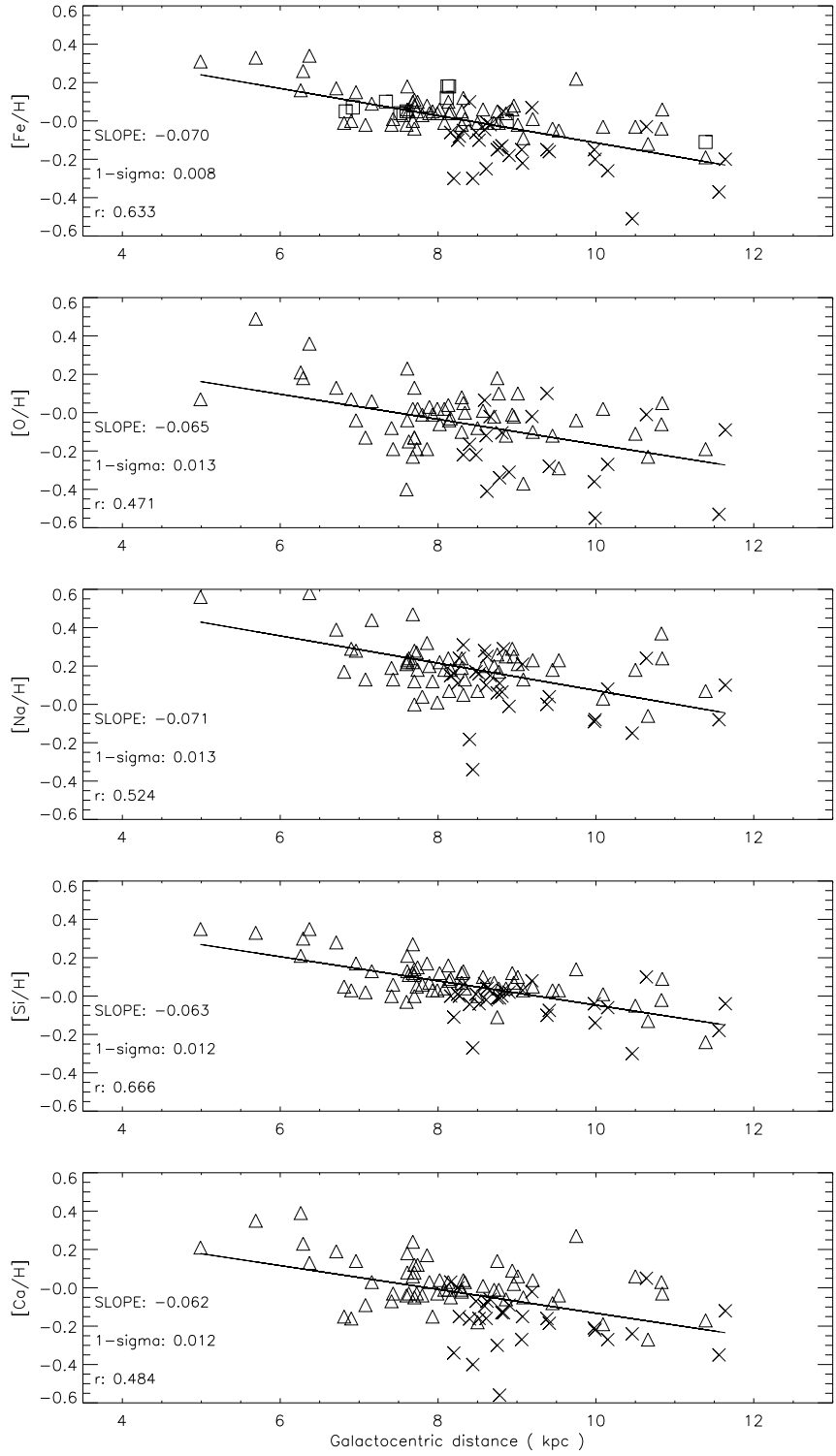}
\caption{Same as Fig. 8, but for Cepheids with Galactocentric distances ranging 
from 5 to 12 kpc.}
\label{fig:whole_data}
\end{center}
\end{figure*}

Futhermore, we also decided to check whether current slope evaluations   
are affected by the Galactocentric radial distances covered by current 
Cepheid sample. Therefore, we included the entire Andrievsky sample (58 Cepheids) 
and the entire \citet{Mot07} sample (11 Cepheids). This means that we increased 
the size of the sample by more than a factor of three and by more than 40\% 
the range of Galactocentric distances. The results of the linear regression 
over the entire sample are listed in Table \ref{whole_grad}. 
The top panel of Fig. \ref{fig:whole_data} once again shows that the slope 
based on the entire data sample agree, within the errors, with those 
based on a smaller Cepheid sample. However, the $1-\sigma$ error is now a 
factor of two lower (0.008 vs 0.19); and the correlation coefficient increased 
by almost the 50\%. These findings bring out several circumstantial evidence: 
{\em i) --} the slope of the iron radial gradient appears marginally affected 
by the uncertainties in radial distances, and indeed the current spread could be 
due to uncertainties in iron abundances;\\
{\em ii) --} The correlation coefficient depends on the intrinsic accuracy of 
individual distances, on the range of Galactocentric distances covered by the 
sample, and to a lesser extent on the spread in metal abundances.\\   
{\em iii) --} The slope based over the entire Cepheid sample is marginally 
affected by the few objects located toward the Galactric centre ($d \le 6$ Kpc)  
and toward the outer reaches of the Galactic disc ($d \ge 11$ Kpc. In fact, 
we performed a linear regression by removing these objects and both the slope 
and the  $1-\sigma$ error are marginally affected.\\  
{\em iv) --} The current iron radial gradient agrees quite well with previous 
determinations. In particular, \citet{Twa97} computed a single-zone gradient 
using open clusters covering a broad radial interval from 6 to 16 kpc, and they 
found -0.067 dex kpc$^{-1}$. The gradient changes to -0.071 dex kpc$^{-1}$ if the cluster 
BE 21 is included in the sample with a high metallicity value and to -0.076 dex kpc$^{-1}$
if it is included with a low metallicity value. A very similar value (-0.06 dex kpc$^{-1}$) 
was also found by \citet{Frie02} using the same chemical composition tracers. By adopting 
classical Cepheids, \citet{Cap01} found a slope of -0.05 dex kpc$^{-1}$, which is 
slightly smaller than the current estimate, but still within the error bars.  
Our result is also fully consistent with one of the last gradient values determined 
with Cepheids by \citet{Kov05}, who added data for 16 distant Cepheids to the Andrievsky's 
sample. By choosing a single-zone gradient in the 4-15.5 kpc range, they obtained a gradient 
of -0.06 dex kpc$^{-1}$. Finally, we can compare our slope with the very recent determination 
by \citet{Luck06}. They found an iron gradient of -0.068 dex kpc$^{-1}$, a value that is in 
remarkable agreement with current estimate. The radial gradients based either on classical 
Cepheids or on open clusters disagree only with the estimates provided by \citet{Kil94}, 
who found a gradient of -0.003$\pm$0.020 dex kpc$^{-1}$, using B-type stars in the 6-15kpc 
region. 

\begin {table*}[!ht]

\scriptsize
\caption {Galactic radial abundance gradient in the 8-12 kpc region. The left part of the tables are the values for our sample, the right part includes
data from \citet{And02a} and subsequent papers and from \citet{Mot07}.}

\begin {tabular}{l|ccc|ccc}
\hline
\hline
Element &     slope      &   1-$\sigma$   & Correlation coeff. &     slope      &   1-$\sigma$   & Correlation coeff. \\
        & dex kpc$^{-1}$ & dex kpc$^{-1}$ &                    & dex kpc$^{-1}$ & dex kpc$^{-1}$ &                    \\
\hline							                                                              
   Fe   &     -0.061     &     0.019      &  0.441             &     -0.056     &     0.012      &  0.410             \\
    O   &     -0.041     &     0.034      &  0.226             &     -0.051     &     0.022      &  0.305             \\
   Na   &     -0.042     &     0.029      &  0.269             &     -0.027     &     0.020      &  0.199             \\
   Si   &     -0.031     &     0.029      &  0.319             &     -0.044     &     0.020      &  0.433             \\
   Ca   &     -0.014     &     0.029      &  0.101             &     -0.024     &     0.020      &  0.160             \\
\hline
     					    
\end {tabular}					       		    
\label {grad}					       	    
\end {table*}	

\begin {table*}[!ht]
\scriptsize
\caption {Galactocentric distances for Cepheids in \citet{And02a} and following
(3 first couples of columns) and in \citet{Mot07} (last couple of columns) 
for which distances based on IR photometry are available.} 

\begin {tabular}{rcrcrc|rc}
\hline
\hline
  Object  & Distance &  Object  & Distance &  Object  & Distance &  Object  & Distance \\
          &   kpc    &          &   kpc    &          &   kpc    &          &   kpc    \\
\hline								                       
   SZ Aql &   6.96   &    X Lac &   9.01   &   AV Sgr &   6.37   &    U Car &   8.11   \\	
   TT Aql &   7.7    &    Y Lac &   9.08   &   BB Sgr &   7.7    &   WZ Car &   8.14   \\	
   FM Aql &   7.86   &    Z Lac &   9.2    & V350 Sgr &   7.61   &   VW Cen &   6.83   \\
   FN Aql &   7.41   &   BG Lac &   8.77   &   RV Sco &   7.74   &   XX Cen &   7.52   \\
 V496 Aql &   7.63   &    T Mon &   9.75   &   RY Sco &   7.16   &   KN Cen &   6.92   \\
 V1162Aql &   7.43   &   SV Mon &  10.83   &   KQ Sco &   6.26   &   GH Lup &   7.56   \\
  Eta Aql &   8.3    &   CV Mon &  10.09   & V500 Sco &   7.08   &    S Mus &   8.12   \\
 V340 Ara &   4.99   &    S Nor &   7.73   &   SS Sct &   7.61   &   UU Mus &   7.6    \\
   RT Aur &   8.94   & V340 Nor &   6.9    &   UZ Sct &   5.69   &    U Nor &   7.34   \\
   SU Cas &   8.71   &    Y Oph &   7.93   &   EW Sct &   8.14   &   LS Pup &  11.39   \\
    V Cen &   7.99   &   BF Oph &   7.68   & V367 Sct &   6.81   &   EU Tau &   8.87   \\
  Del Cep &   8.57   &   GQ Ori &  10.84   &   BQ Ser &   7.7    &          &          \\	
   BG Cru &   8.3    &    X Pup &  10.5    &   ST Tau &   9.53   &          &          \\
    X Cyg &   8.32   &    S Sge &   8.13   &   SZ Tau &   8.96   &          &          \\
   VZ Cyg &   8.75   &    U Sgr &   7.89   &    S Vul &   7.6    &          &          \\	
 Beta Dor &   8.5    &    W Sgr &   8.08   &    T Vul &   8.34   &          &          \\
    W Gem &   9.45   &    Y Sgr &   8.02   &    U Vul &   8.16   &          &          \\
   RZ Gem &  10.66   &   VY Sgr &   6.29   &   SV Vul &   7.8    &          &          \\	
   AD Gem &  11.39   &   WZ Sgr &   6.71   &          &          &          &          \\
 Zeta Gem &   8.86   &   AP Sgr &   7.68   &          &          &          &          \\
\hline
     					    
\end {tabular}					       		    
\label {adddist}					       	    
\end {table*}

\begin {table*}[!ht]

\scriptsize
\caption {Galactic radial abundance gradient in the 5-12 kpc region. Our sample has been completed
with Cepheids from \citet{And02a} and following and from \citet{Mot07} for which a homogeneous 
determination of NIR distances was possible.}

\begin {tabular}{l|ccc|ccc}
\hline
\hline
Element &     slope      &   1-$\sigma$   & Correlation coeff. \\
        & dex kpc$^{-1}$ & dex kpc$^{-1}$ &                    \\
\hline							       
   Fe   &     -0.070     &     0.008      &  0.633             \\
    O   &     -0.065     &     0.013      &  0.471             \\
   Na   &     -0.071     &     0.013      &  0.524             \\
   Si   &     -0.063     &     0.012      &  0.666             \\
   Ca   &     -0.062     &     0.012      &  0.484             \\
\hline
     					    
\end {tabular}					       		    
\label {whole_grad}					       	    
\end {table*}	

\subsection{Galactic radial gradient for other elements} 

Galactic radial abundances gradients for the other elements in the 8-12 kpc region, 
computed either from current sample or from the sample implemented with Cepheids from 
the Andrievsky's sample are once again in good agreement. The discrepancies between 
the two different slopes are on average of the order of 0.01 dex kpc$^{-1}$. The only 
exception is for $Na$ and $O$, for which the discrepancy is $\sim 0.015$ dex kpc$^{-1}$, 
but still within the error bars.

The radial gradients for the other elements in the radial range from 8 to 12 kpc
present larger uncertainties when compared with the iron one. A similar trend was 
also found by \citet{And02a}.   
Once again, current slopes are higher when compared with those estimated by 
\citet{And02a}. In particular, for oxygen they found -0.022$\pm$0.009 dex kpc$^{-1}$, 
while we find a steeper slope -0.041$\pm$0.034 dex kpc$^{-1}$.
The same outcome applies to $Na$ for which they found -0.023$\pm$0.006 dex kpc$^{-1}$, 
while we find -0.042$\pm$0.029 dex kpc$^{-1}$. On the other hand, the slopes for 
$Si$ and $Ca$ agree very well, and indeed the slopes are 
-0.030$\pm$0.004 dex kpc$^{-1}$ and -0.021$\pm$0.006 dex kpc$^{-1}$, while we 
find -0.031$\pm$0.029 dex kpc$^{-1}$ and -0.014$\pm$0.029 dex kpc$^{-1}$.

If we compute the radial gradients using the entire Cepheid sample, they 
appear to be similar to the iron gradient and range from -0.06 to 
-0.07 dex kpc$^{-1}$. The gradient appears to be flatter across the solar 
circle.  This behaviour was expected, because it has already been detected, first by
\citet{Twa97} but also by \citet{And02a}. This circumstantial evidence supported the use of a 
multi-zonal gradient. However, before firm conclusions can be derived, a large sample 
of homogenous chemical abundances for iron and heavy elements is required. 

As far as oxygen is concerned, the comparison with previous studies shows a 
good agreement with \citet{Sma97}, \citet{Gum98}, and \citet{Roll00}, who found 
slopes of -0.07$\pm$0.01 dex kpc$^{-1}$, -0.067$\pm$0.024 dex kpc$^{-1}$, 
and -0.067$\pm$0.008 dex kpc$^{-1}$, using B-type stars as chemical composition 
tracers. Futhermore, \citet{Mac98} found -0.058$\pm$0.007 dex kpc$^{-1}$ using 
planetary nebulae and \citet{Aff97} found~-0.064$\pm$0.009 dex kpc$^{-1}$ from 
$HII$ regions. However, current results do not support the flat slopes for the 
oxygen radial gradient found by \citet{Fitz92} (-0.03$\pm$0.02 dex kpc$^{-1}$) 
and \citet{Kil94} (-0.021$\pm$0.012 dex kpc$^{-1}$) using B-type stars or
by \citet{Deh00} (-0.0391$\pm$0.005 dex kpc$^{-1}$) using $HII$ regions. 
Obviously they also do not agree with the null radial gradient found by 
\citet{Kau94} (-0.000$\pm$0.009 dex kpc$^{-1}$) using B-type stars.

For silicon, our results are in good agreement with \citet{Roll00}, who found a 
slope of -0.06$\pm$0.01 dex kpc$^{-1}$, but agree neither with the flatter 
slopes by \citet{Kil94} (+0.000$\pm$0.018 dex kpc$^{-1}$)  nor 
with the steeper slopes by \citep{Gum98} (-0.107$\pm$0.028 dex kpc$^{-1}$).
Both these authors were using B-type stars.

However, in order to ascertain whether the difference in the different slope 
estimates are dominated by deceptive empirical uncertainties or they might 
be intrinsic, a detailed investigation of stellar ages and 
radial distribution of the different chemical tracers is required.

\section{Conclusions}

We have determined the abundances of iron and six light metals (O, Na, Mg, Al, Si, Ca) in 
30 Galactic Cepheids. These abundances were used to compute a slope for the Galactic 
radial abundance gradient. For the iron Galactic gradient in the radial distance range 
from 8 to 12 Kpc we found slopes of -0.061 dex kpc$^{-1}$ and of -0.056 dex kpc$^{-1}$ 
if we include 69 Cepheids with homogeneous distance determinations. The current slope is 
steeper when compared with the slope estimated by \citet{And02a} for the central region 
of the multi-zonal gradient. On the other hand, the current gradient computed over a more 
extended baseline (-0.070$\pm$0.008 dex kpc$^{-1}$) is fully consistent with other 
studies based either on open clusters or on Cepheids. Current findings do not allow us 
to provide firm constraints on the multi-modal model, due to the limited number of 
targets in the crucial zone. The hypothesis of a multi-modal gradient deserves 
further investigations, mainly to fill the transition region across 10-11 kpc.

%

\begin{acknowledgements}
This research made use of the SIMBAD and VIZIER databases, operated at the CDS, 
Strasbourg, France. We would also to thank A.M. Piersimoni for help in estimating
the mean NIR magnitudes based on 2MASS photometry. We deeply thank
F. Pont and F. Kienzle who did the observations.   
\end{acknowledgements}

\bibliographystyle{aa}

\end{document}